\documentclass[final]{ustcstep}

\usepackage{lineno}
\usepackage{graphicx}
\usepackage{lscape}
\usepackage{multirow}
\usepackage{color}
\usepackage{url} 

\newcommand\addr[2]{{\footnotesize \it $^{#1}$#2}\\}
\usepackage[pdfborder={0 0 0},urlcolor=blue,breaklinks]{hyperref}

\usepackage{enumerate}
\usepackage{natbib}

\begin{document}

\title{Satistical Study of the Interplanetary Coronal Mass Ejections from 1996 to 2014}

\author{Yutian Chi$^{1,2}$, Chenglong Shen$^{1,3,*}$, Yuming Wang$^{1,4}$, Pinzhong Ye$^{1}$, and S. Wang$^{1,5}$\\
  \addr{}{$^1$ CAS Key Laboratory of Geospace Environment, Department of Geophysics and Planetary Sciences, }\\
  	\addr{}{University of Science \& Technology of China, Hefei, Anhui 230026, China (clshen@ustc.edu.cn)}\\
	 \addr{}{$^2$ Sate Key Laboratory of Space Weather, Chinese Academy of Sciences, Beijing 100190}\\
	 \addr{}{$^3$ Synergetic Innovation Center of Quantum Information \& Quantum Physics,}\\
	 \addr{}{ University of Science and Technology of China, Hefei, Anhui 230026, China}\\
	 \addr{}{$^4$ Collaborative Innovation Center of Astronautical Science and Technology, China}\\ 
	 \addr{}{$^5$ Mengcheng National Geophysical Observatory, School of Earth \& Space Sciences,}\\
	 \addr{}{ University of Science \& Technology of China, Hefei, China}\\
	 \addr{}{$^*$ Corresponding author}}

\maketitle
\tableofcontents

\begin{abstract}
In this work, we establish an ICME list from 1996 to 2014 based on the in-situ observations from the WIND and ACE satellites. 
Based on this ICME list, we extend the statistical analysis of the ICMEs to the solar maximum phase of solar cycle 24th. 
The analysis of the annual variations of the properties of ICMEs show that the number of ICMEs, the number of shocks, the percentage of ICMEs drove shocks, 
the magnetic field and plasma properties of ICMEs are well correlated with the solar cycle variation.  
The number of MCs do not show any correlation with sunspot number. But, the percentage of the MCs in ICMEs show good anti-correlation with the sunspot number.
By comparison the parameters of MCs with None-MC ICMEs, we found that the MCs are stronger than the None-MC ICMEs .
In addition, we compare the parameters of ICMEs with and without shocks. It is found that the ICMEs with shocks are much stronger than the ICME without shocks.  
Meanwhile, we discuss the distribution of the magnetic field and solar wind plasmas parameters of the sheath regions of ICMEs at first time.
 We find that the magnetic field and solar wind velocity in the sheath region are higher than them in the ejecta of ICMEs from statistical point of view.
\end{abstract}

ection{Introduction}

Interplanetary Coronal Mass Ejections (ICMEs) are the interplanetary counterparts of the Coronal Mass Ejections (CMEs).
From 1970s, the ICMEs has been reported and studied by the in-situ measurements for decades\citep[e.g.][]{Gosling1973,Burlaga1981}. 
Based on the literatures, the possible in-situ signatures of the ICMEs are the strong magnetic field, rotated magnetic field, low plasma $\beta$,
low proton temperature, expansion velocity profile, bidirectional electron streaming, lower energetic particle intensity, abnormal charge state of ions, high total pressures, 
and so on\citep[e.g.] [and reference therein]{2001JGR...10620957B,Cane2003d,Jian2006c,2006SSRv..123..177W,2006SSRv..124..145G,Kilpua2009,Richardson:2010jq,Jian2011,Kilpua2012,Kilpua2014a}.  
It should be noted that, none of these signatures can be observed by all ICMEs. 
Thus, different authors always used different criterion to identify the ICMEs from in-situ observations.
Using different criterion, different lists about the ICMEs and their related structures (such as ICME like structures and so on) 
were compiled by different authors\citep[e.g.] [and reference therein]{Jian2006c,Lepping2006,Richardson:2010jq,Kilpua2012,Wu2015}. 

The magnetic cloud (MC), which was thought to be a special type of ICMEs, was first reported by \citet{Burlaga1981}.  
The signatures of the MCs are the enhanced magnetic field strength, long and smooth rotation of the magnetic field vector and low proton temperature.
Differ from the idenfication of ICMEs, a structure, which called as MC, have to fit at least all these three signatures.
To study the signature of the MC, different lists of MCs have also been established\citep[e.g.][]{Lepping2006,Wu2015}. 
Meanwhile, in the ICME lists from Lan Jian\citep{Jian2006c} and Richardson \& Cane\citep{Richardson:2010jq}, they also show the result that whether a ICME is an MC. 
Based on these lists and other observations and different models, the parameters of MCs and the comparison between the MC and normal ICMEs has been discussed\citep[e.g.][and reference therein]{Jian2006c,Lepping2006,Lepping2014,Wu2015,Wang2015}.

In this work, we will establish a new online ICME catalogue mainly based on the WIND magnetic field and plasma observations from 1996 to 2014. 
Meanwhile, the suprathermal electron pitch angle distribution observations from wind and ACE satellites and the proton and electron flux observation from WIND are also included.
Based on this catalogue, we extend the statistical analysis of ICMEs and their related structures to the end of 2014. 
The method we use to determine the ICMEs and the brief introduction of the catalogue are shown in Section 2. 
In Section 3, we will discuss the variation of the annual number of ICMEs.
Meanwhile, the annual numbers and ratios of MCs and shocks driven by these ICMEs will also be discussed in this section.
 Further, the properties of ICMEs and their annual variation will be shown in Section 4. 
In Section 5, we will discuss the properties of the ICMEs with shocks and compare them with ICMEs without shocks. 
In addition, the properties of the sheath regions of the shocks driven by ICMEs will be shown in Section 5. 
At last, a briefly summary will be given in the last section.

\section{ICME Catalogue}\label{sect2}

\begin{figure}
	\center
	\noindent\includegraphics[width=0.7\hsize]{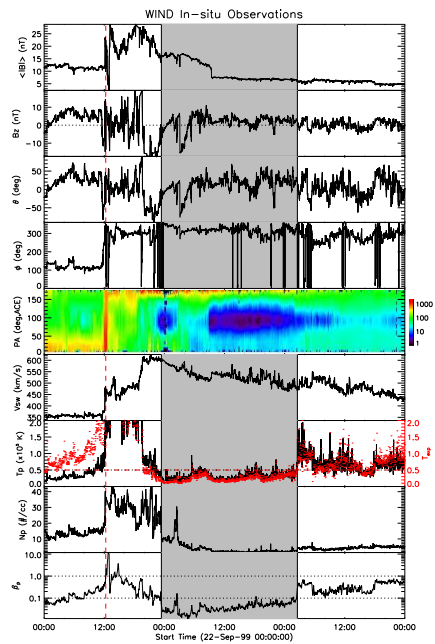}
	\caption{An example of the interplanetary coronal mass ejection (ICME): the 22 - 24 September 1999 event.  From top to the bottom, they are the magnetic field
		strength ($|B|$), the z-component of the magnetic field in Geocentric Solar Ecliptic (GSE) coordinate system ($B_z$), the elevation ($\theta$) and azimuthal ($\phi$) of
		field direction in GSE coordinate system from WIND, the suprathermal electron pitch angle distribution from ACE, solar wind speed ($V_{SW}$), 
		proton density ($N_p$), proton temperature ($T_p$) and the ratio of proton thermal pressure to magnetic pressure ($\beta_p$) from WIND observations.}
	\label{icme}
\end{figure}

In the literatures, different signatures were used to identify the ICMEs. 
In this work, the criterion we used are: (1) enhanced magnetic field intensity, (2) smoothly changing field direction, (3) declining profile of the solar wind velocity,
(4) relatively low proton temperature, (5) low proton plasma beta, and (6) bidirectional streaming of electrons.
A structure is recognized as an ICME when it fits at least three of the criterion listed above as the same as we did in \citet{Shen2014a}.
Figure \ref{icme} shows an example of the ICME: the 22 - 24 September 1999 event. From 23:24UT 22 September 1999 to 02:33UT 24 September 1999 (as the gray region shown), 
the in-situ observations show signatures with expansion velocity, low temperature, low plasma beta $\beta$ and obvious bi-direction electron streaming. 
Thus, this is an ICME structure which fit at least 4 criterion we used. About 11 hours early, the shock drive by this ICME was detected by WIND (as the red vertical dashed line shown). 
After the shock, the magnetic field strength, velocity, density, temperature enhanced significantly.
It should be noted that, for some events, the boundaries of the ICMEs are hard to be determined based on the magnetic field and plasma observations only. 
For these events, the energetic particle signatures are also included to check the boundaries of the ICME structures\cite[e.g.][and reference therein]{Cane:2006kg}.

\begin{figure}
	\center
	\noindent\includegraphics[width=0.7\hsize]{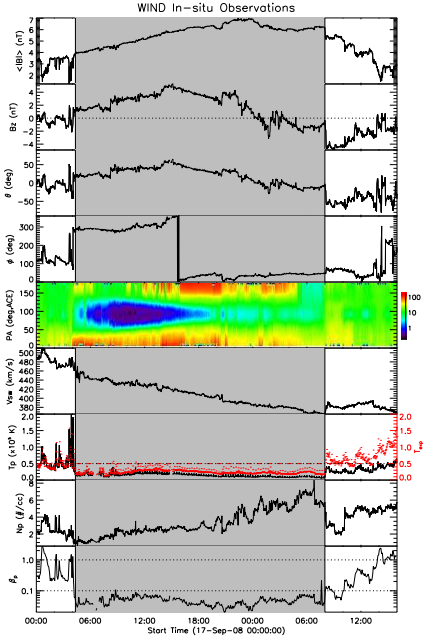}
	\caption{An example of the magnetic cloud (MC) event: 17 - 18 September 2008 event.}
	\label{mc}
\end{figure}

Based on these criterion, we totally identified 436 ICME events from 1996 to 2014, and  about half (217) of them drove shocks. 
In addition, we further make a judgement about whether an ICME is a MC using the criterion of obvious enhanced magnetic field, clear and smooth rotation of the magnetic field vector
 and the low plasma beta.
Figure \ref{mc} shows an example of the MC: 17 - 18 September 2008 event. This events has been studied by \cite{Wang2014b}. 
Seen from this figure, during the period of 04:20 UT 17 September to 08:00 UT 18 September (as the gray region shown), all signatures of MC are clear. 
In addition, the other signatures of a typical ICME or MC structure, such as declining profile of the solar wind velocity and the bidirectional streaming of suprathermal electrons, are also clear. 
Based on these criterion, we totally found 146 MC in our ICME list. Thus, there are about 33\% ICMEs are MCs,  such ratio is as the similar as \citet{Wu2010} and \citet{Richardson:2010jq}.

\begin{figure}
	\center
	\noindent\includegraphics[width=\hsize]{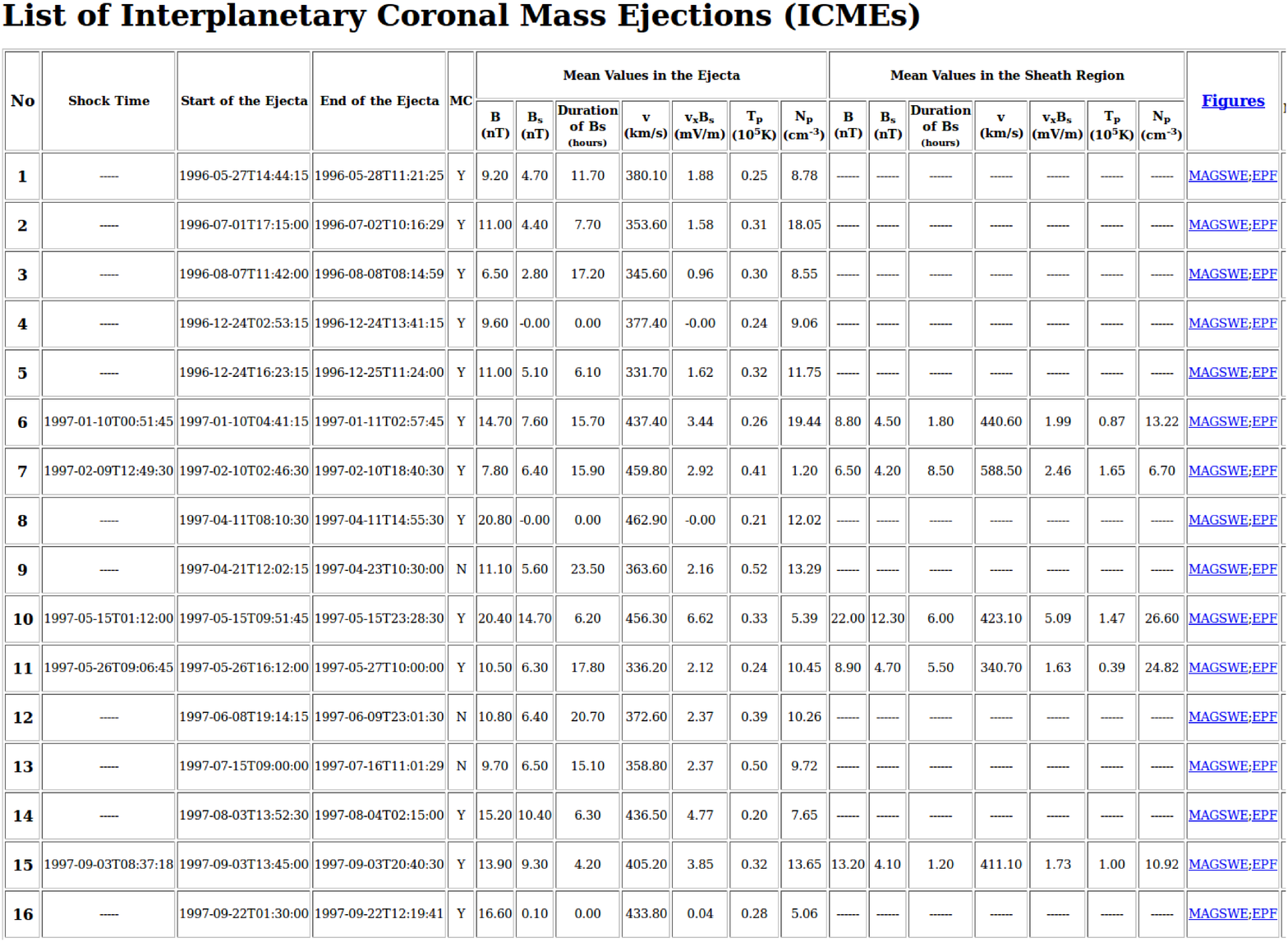}
	\caption{A snapshot of the online ICME catalogue at \url{http://space.ustc.edu.cn/dreams/wind_icmes/}.}
	\label{cata}
\end{figure}

An online catalogue of the ICMEs locates at \url{http://space.ustc.edu.cn/dreams/wind_icmes/}. Figure \ref{cata} shows a snapshot of the online catalogue.
The 1st column in the table shows the order numbers of the ICME events.
The 2nd column lists the times of the arrival of the shocks driven by these ICMEs. The symbol `---' in this column indicates that no shock driven by the ICME could be found.
The beginning and the end time of the ICMEs are shown in the 3th and 4th columns. 
The 5th column gives the result that whether this ICME is an MC. 
The symbols `Y' indicate that these ICME are MCs while symbols `N' mean that these events are not MC.
The 6th to 12th columns gives the values of the $B$, $B_s$, duration of $B_s$ ($\Delta t$), $v$, $v_xB_s$, $T_p$ and $N_p$ of these ICMEs's ejecta region, 
and the 13rd to 19th columns show these parameters for the sheath regions driven by them. 
In these columns, the mean values of the magnetic field and solar wind plasma parameters during the ejecta and the sheath regions are used.
If there was no shock driven by a ICME, the values in columns 13rd to 19th are `---'. 
In addition, there are large magnetic field data gap for 1 event and solar wind plasma data gap for 12 events in our list. 
For these events, we do not calculate the parameters of them and use the string of  `Poor data' in the column 6th to 19th to show them.
The links of the magnetic field and solar wind plasma images of these ICMEs, as similar as Figure \ref{icme} shown, could be found at the 20th column in the online catalogue with the symbols of `MAGSWE'.
Meanwhile, we also provide the electron and proton flux images from WIND/3DP in the online catalogue via the link on the symbols of `EPF' in column 20th. 

In recent, there are some other ICMEs catalogues. 
Two of them are the (1) Richardson \& Cane's catalogue (RC catalogue, \citep{Richardson:2010jq}), (2) Lan Jian's catalogue ( JL catalogue \citep{Jian2006c}).
The time period of RC catalogue is almost the same with us. By compared our list with their, we found that large fraction of ICMEs in these two catalogues are same. 
About 81\% (356) ICMEs in our catalogue are also listed in their list. Meanwhile, there are 81 events only listed in our catalogue and 113 events only listed in CR catalogue.
Meanwhile, the time period of JL catalogue is from 1995 and only up to 2009. 
During same period from 1996 to 2009, we totally identified 248 ICME events and 71\% (176) of them are also listed in JL catalogue. 
These comparison show that these catalogues are similar for most cases but show difference.
The main reason is that we use different criterion to identify the ICMEs from in-situ observations. 
In our catalogue, we mainly consider the magnetic field, velocity, temperature, plasma beta observations.
In addition, we take the solar wind electron pitch-angle observations from ACE and the proton and electron fluxes observations from WIND into account.
The pitch-angle observations are used to identify the possible ICMEs and the proton and electron fluxes observations are used to determine the possible boundaries of the ICMEs.
Differ from us, in RC catalogue, the ion composition and charge state observations are taken in to consideration. 
Meanwhile, the ratio of alpha particles to protons and the total perpendicular pressure are used in JL catalogue.

\section{Annual Numbers of ICMEs}

\begin{figure}
	\center
	\noindent\includegraphics[width=0.9\hsize]{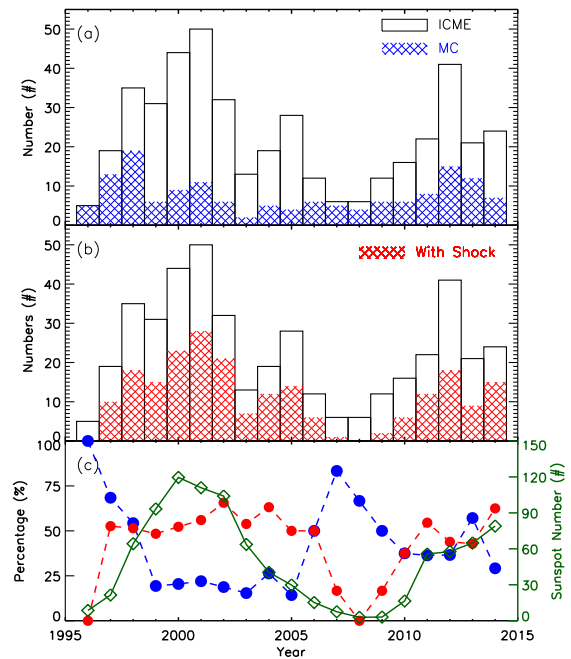}
	\caption{The annual numbers of the ICMEs, MCs and the shocks. Panel (a) shows the annual numbers the total ICMEs and the MCs (blue). The Panel (b) shows
		the annual numbers the total ICMEs and the shocks (red). Panel (c) shows the ratios of the MCs (blue solid circles) and the ICMEs with shocks (red solid cycles) varied with time. The dark green diamonds shows the annual mean sunspot numbers varied with time.}
	\label{year_icme_number}
\end{figure}

Panel (a) and (b) in Figure \ref{year_icme_number} show the annual numbers of ICMsE, MCs (blue bars in panel (a)) and ICMEs with shocks (red bars in panel (b)) from 1996 to 2014. 
Seen from these panels, the annual numbers of ICMEs and the numbers of the shocks driven by the ICMEs are well correlated with the solar cycle variation.
The correlation coefficient between the annual ICME numbers and the sunspot numbers is 0.83, 
while the correlation coefficient between the numbers of the shocks driven by the ICMEs and the sunspot numbers is 0.88.
Both the peak annual numbers of ICMEs and shocks driven by ICMEs are located at 2001, with the maximum values of 50 and 28 respectively.
These correlations could be understood by the fact that the numbers of CMEs and also the fast ones which can drive shocks are correlated with the solar cycle variation\citep[e.g.][and reference therein]{2003ESASP.535..403G,Wang2014}.
The red solid circles in Panel (c) show the percentages of ICMEs which drove shocks in each year while the dark green diamonds show the annual sunspot numbers. 
Seen from this panel, the percentages of ICMEs drove shocks are also correlated with the sunspot numbers. The correlation coefficient between them is 0.67. 
These results indicate that not only the numbers of ICMEs and shocks driven by them but also the possibilities of the ICMEs drove shocks are higher in solar maximum.

Meanwhile, seen from Panel (a) and (c), the annual numbers of MCs are not correlated with the sunspot numbers, which has also been suggest by \cite{Lepping2014}.
The peak annual number of MCs is found at 1998 with the value of 19. 
The blue solid circles in Panel (c) show the percentage of MCs in all ICMEs for each year. 
Different from the shocks driven by ICMEs, the percentages of the MCs show obvious anti-correlation with the sunspot numbers. 
Large faction (almost all in 1996) of ICMEs are magnetic cloud in solar minimum but only about 20\% ICMEs are MCs in solar maximum.
The correlation coefficient between the percentages of the MCs and the sunspot numbers is -0.66. 
It further confirms the results that the ICMEs are more like to be MCs in solar minimum\citep[e.g.][]{Cane2003d,Richardson2004a,Jian2006c,Wu2010,Richardson:2010jq}
One possible reason is that the Coronal Mass Ejections (CMEs) are more likely to deflect the solar equator during their propagation in the corona in the solar minimum
\citep[e.g][and reference therein]{Cremades:2004cz,Shen2011,Gui:2011dw}. 
Thus, the satellite located near the ecliptic plane are more likely to detect the `core' regions of CMEs which always treated as the MCs.

\section{Statistical Analysis of the Properties of ICMEs}

\subsection{Distribution of the Magnetic Field and Plasma Parameters}

\begin{figure}
        \center
        \noindent\includegraphics[width=\hsize]{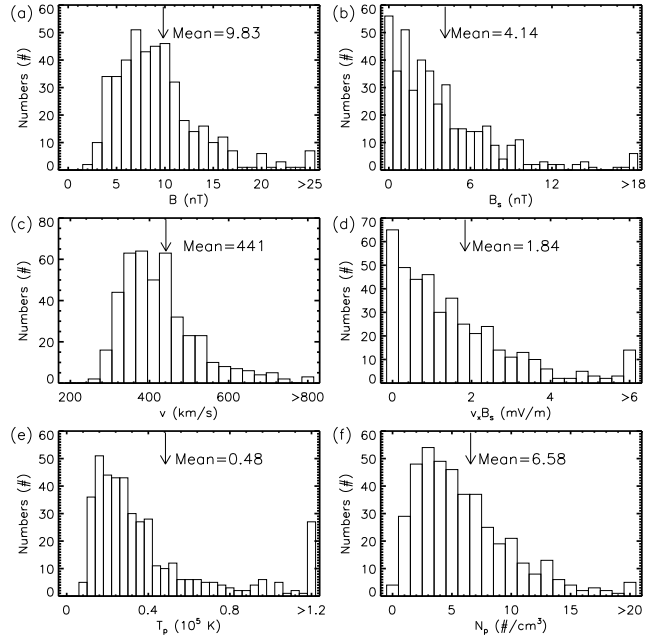}
        \caption{The distributions of different parameters of ejecta of ICMEs. Panel (a) - (e) show the distribution of $B$, $B_s$, $v$, $v_xB_s$, $T_p$ and $N_p$ respectively. 
The arrows show the mean values for all ICMEs.
}
        \label{hist}
\end{figure}

Figure \ref{hist} shows the distributions of magnetic field and  plasma parameters of the ejecta of ICMEs. 
For each ICME event, the mean values of these parameters during its pass through the satellite are used.
Panel (a) - (f) show the distributions of the magnetic field strength ($B$), the south component of the magnetic field ($B_s$), the velocity ($v$), the dawn-dust electric field ($v_xB_s$), 
the proton temperature ($T_p$) and the proton number density ($N_p$) respectively. 
The mean values of these parameters for all ICMEs are indicated by the arrows in each panels.
Based on this figure, we found that:

\begin{enumerate}

\item As the Panel (a) shown, the distribution of the magnetic field strength is wide, from about 2.80 nT to more than 37.0 nT. 
The mean values for all the ICMEs is 9.83 nT, about twice of the background magnetic field (5 nT) in the solar wind. 
This value is similar but little bit smaller than the value of 10.1 obtained by \citet{Richardson:2010jq}.

\item The velocities of the ICMEs concentrate to the value of 450 km/s as Panel (c) shown. The velocities of 78\% ICMEs located in a narrow range from 300 km/s to 500 km/s.
This might be caused by the acceleration (or deceleration) of CMEs controlled by the solar wind during its propagation from Sun to Earth\citep[e.g.][and reference therein]{Gopalswamy2000,Lugaz:2012es,Vrsnak2012}. 
The mean value of the velocity is 441 km/s, which is also smaller than the value (476 km/s) obtained by \citet{Richardson:2010jq}.

\item The south component of the magnetic field and the dawn-dusk electric field are important factors in determine the geoeffectiveness of ICMEs\citep[e.g.][]{Gonzalez1994}. 
Seen from Panel (b) and (d), the south component of the magnetic field and the dawn-dusk electric field $v_xB_s$ also varied in a large range. 
The mean and the maximum value of the $B_s$ are 4.14 nT and 35.4 nT, while the mean and maximum values of the $v_xB_s$ are 1.83 mV/m and 24.2 mV/m.  
In addition, only about 14\% and 15\% of the ICMEs with $B_s$ and $v_xB_s$ close to 0. 
That means most of ICMEs carried south component of the magnetic field and possible to cause geomagnetic storms.

\item Panel (e) and (f) shows the distribution of the proton temperature and the proton number density. The mean values of these two parameters are 4.84$\times$10$^5$ K and 6.58 cm$^{-3}$.
They are also similar but little bit smaller than the previous results obtained by \cite{Richardson:2010jq}.
\end{enumerate}

\subsection{Comparison between MC and Non-MC ICMEs}
\begin{figure}
        \center
        \noindent\includegraphics[width=\hsize]{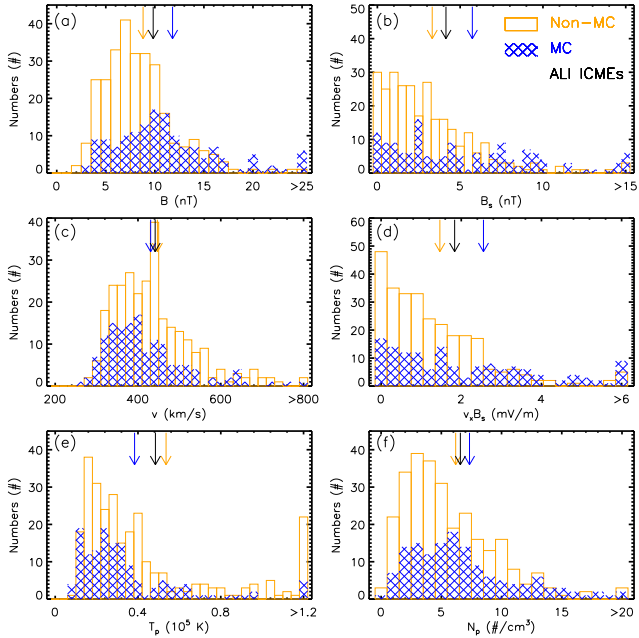}
        \caption{The distributions of the magnetic field and plasma parameters of the ejecta of MCs and Non-MC ICMEs. 
	The arrows show the mean values. The blue and orange colors show the MC and Non-MC ICMEs events respectively. }
        \label{compare_mc}
\end{figure}

MC is a special type a ICME. 
Figure \ref{compare_mc} show the distribution of the magnetic field and plasma parameters for ejecta of the MCs and non-MC ICMEs.
As the same as the previous results shown, the magnetic field related parameters ($B$, $B_s$ and $v_xB_s$) for the ejecta of the MCs and Non-MC ICMEs are significant different. 
For the MCs, these parameters are larger than the values of the Non-MC ICMEs. 
The mean values of $B$ for MCs and Non-MC ICMEs are 11.8 nT and 8.80 nT respectively, which is as similar as \cite{Richardson:2010jq} obtained.
The mean values of the $B_s$ and $v_xB_s$ for the MCs are 5.74 nT ad 2.55 mv/m. For Non-MC ICMEs, they are much smaller with the values of 3.31 nT and 1.47 mV/m.
Shown in Panel (c), the velocity distribution of the MCs and the Non-MC ICMEs are almost the same, and, the mean values (430 km/s for MCs and 447 km/s for Non-MC ICMEs) 
are all close to the background solar wind speed.
This indicates that the dynamic evolution of MC and the Non-MC ICMEs during their propagation in interplanetary space are all controlled by the background solar wind.
It is obvious that the temperature of the MCs are smaller than the Non-MC ICMEs as Panel (e) in Figure \ref{compare_mc}.
In addition, the proton density did not show any difference between the MC and Non-MC ICMEs as the Panel (f) shown.

\subsection{The Solar Cycle Variation of the Magnetic Field and Plasma Parameters}

\begin{figure}
	\center
	\noindent\includegraphics[width=0.9\hsize]{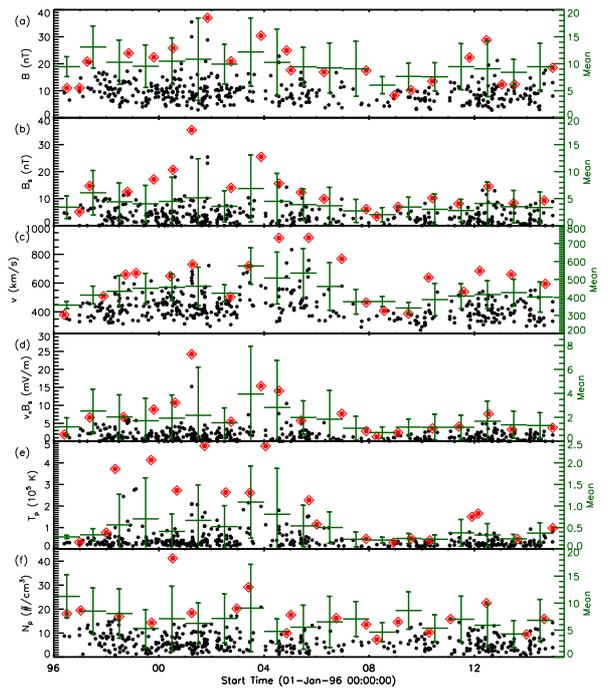}
	\caption{The parameters of the ejecta of ICMEs varied with time. Each dots in Figure \ref{para_year} show an ICME event, and, the red dots in the diamonds show the events with the annual maximum value in each year. The dark green horizontal lines show the annual mean values of the ejecta of ICMEs.}
	\label{para_year}
\end{figure}

Figure \ref{para_year} shows the parameters of ICMEs varied with time from 1996 to 2014. 
Each dot shows an ICME event. The red dots in the diamonds show the events with the annual maximum values in every year.
Seen from panel (a) to (e), the annual maximum values of $B$, $B_s$, $v$, $v_xB_s$ and $T_p$ for the ICMEs well correlated with the solar cycle variation. 
In solar maximum, the annual maximum vales of the ICMEs are larger while in the solar minimum they are smaller. 
The maximum events of the $B$, $B_s$ and $v_xB_s$ from 1996 to 2014 is an MC in 31 March 2001 event (No 141 event in the online catalogue). 
This event has been studied by \citet{Wang2003}. Based on  \citet{Wang2003}, this MC is a part of a multiple ICMEs structure.
Thus, the strong but short duration magnetic field in this MC might be caused by the interaction (or compression) between multiple MCs during their propagation from Sun to Earth.
The ICME with highest CME velocity is the 11 - 12 September 2005 event (No 271 in the catalogue) with $v=916\;km/s$.

In addition, the annual mean values of the magnetic field and plasma parameters (as the dark green horizontal lines shown) are also varied with solar cycle. 
They are higher in solar maximum and lower in solar minimum. Almost all the parameters are highest at the year of 2003, which is later than the peak of the sunspot numbers..
In addition, the solar cycle variation of the velocities and the $v_xB_s$ are most significant.
It should be noted that, from the year of 2008, almost all the parameters of the ICMEs are smaller than before.
Such decreasing of ICME parameters of ICMEs has been reported by \citet{Jian2011} and \citet{Kilpua2012}.

On the contrary, the mean value and the maximum values of the proton density did not show obvious solar cycle variation.

\section{The ICMEs with shocks}

\begin{figure}
	\center
	\noindent\includegraphics[width=\hsize]{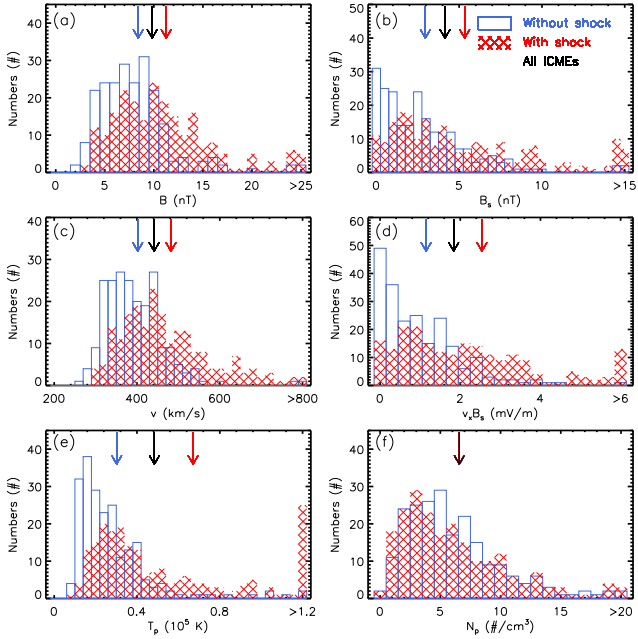}
	\caption{Comparison of the parameters between the ejecta of ICMEs with and without shocks. Panel (a) - (d) show the $B$, $B_s$, $v$, $v_xB_s$, $T_p$ and $N_p$ respectively.}
	\label{compare_shock}
\end{figure}

In Section \ref{sect2}, we found that about half of the ICMEs drove shocks.  
We can expect that the ICMEs with shocks are faster than the ICMEs without shocks, because that only fast CMEs can exceed the local alfv\'en speed and them drove shocks.
Figure \ref{compare_shock} shows the comparison of the parameters between ejecta in the ICMEs with and without shocks.
It is obviously that the distribution of the $B$, and $B_s$, $v$, $v_xB_s$ and $T_p$ are obvious different for the ICMEs with and without shocks.
These results show that the ICMEs with shocks are stronger than the ICMEs without shocks\citep{2006SSRv..124..145G}. 
Seen from Figure \ref{compare_shock} (c), there are some fast CMEs did not drive shocks while some slow CMEs drove shocks. 
One possible reason is the kinematic evolution of the ejecta in the interplanetary space, which might cause the velocities of the ejecta changed greatly during their propagation in interplanetary space\citep[e.g.][and reference therein]{Gopalswamy2000,Lugaz:2012es,Vrsnak2012}. .
 Another possible reason is that the background plasma and magnetic field condition are not the same 
 and then the background alfv\'en speed might vary in a large range in different cases. 
\citet{Shen:2007ww} show two examples that the fast CME drove a weak shock and a slow CME drove a stronger shock near the Sun because of the different background alfv\'en speed.

\begin{figure}
	\center
	\noindent\includegraphics[width=\hsize]{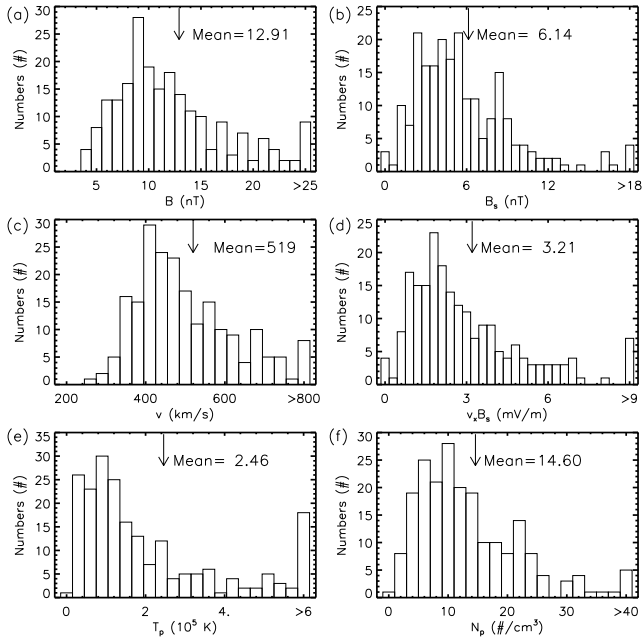}
	\caption{The distributions of different parameters of sheath regions. Panel (a) - (e) show the distribution of $B$, $B_s$, $v$, $v_xB_s$, $T_p$ and $N_p$ respectively. }
	\label{hist_sheath}
\end{figure}

Figure \ref{hist_sheath} shows the distribution of the magnetic field and plasma parameters in the sheath regions. 
Seen from Panel (a) to (f) in this figure, the distribution of all these parameters are similar but larger than the ejecta.
The mean values of the $B$, $B_s$, $v$, $v_xB_s$, $T_p$ and $N_p$ are 12.9 nT, 6.13 nT, 520 km/s, 3.21 mV/m, 2.45$\times$10$^5$ K and 14.6 cm$^{-1}$,
which are higher than them in ejecta.
By checked all the ICMEs with shocks, we found that the magnetic field in the sheath region is stronger than the ejecta for about 60\% events.
Meanwhile, there are 77\% ICMEs  whose velocities in the sheath regions are faster than the ejecta.
Thus, we conclude that the magnetic field and the velocity in the sheath region are higher than them in the ejecta.




\section{The Correlation between Different Parameters}

\begin{figure}
	\center
	\noindent\includegraphics[width=\hsize]{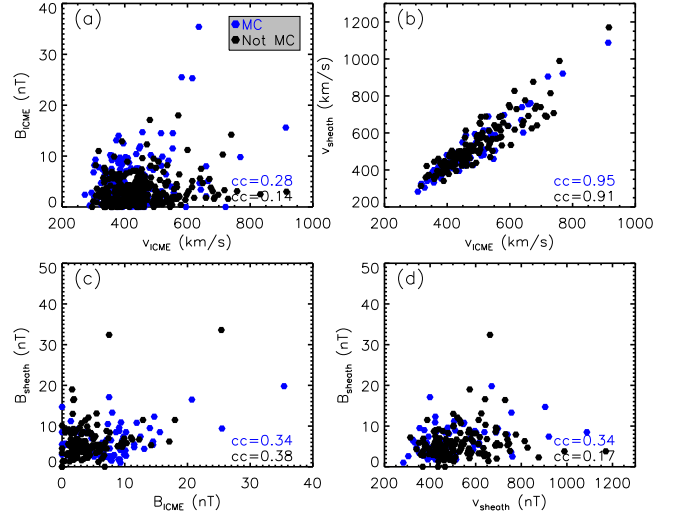}
	\caption{The scatter plots of the magnetic field $B$ and the velocities $v$ in the ejecta and the sheath regions.}
	\label{para_corr}
\end{figure}

Figure \ref{para_corr} shows the scatter plot of the magnetic field strength and the velocities.
 Different panels show the correlation between $B$ and $v$ in the ejacta and sheath regions respectively. 
 Seen from Panel (b), the velocities in the sheath regions and the ejecta are well correlated.
 The correlation coefficient between them are 0.95 for MC and 0.91 for non-MC ICMEs.
 Such good correlation has been reported by \citet{2006SSRv..124..145G}.
 Meanwhile, there are not obvious correlations between other groups of parameters. 
 The correlation coefficients of them are all smaller than 0.4.

\section{Summary}

In this work, we have established an ICME catalogue from 1996 to 2014 mainly based on the magnetic field and plasma observations from WIND.
Meanwhile, the suprathermal electron pitch angle distribution from WIND and ACE satellites and the proton and electron flux observation from WIND are also included.
The link of this catalogue is \url{http://space.ustc.edu.cn/dreams/icmes/}. 
The time of the shock, the time of the beginning and the end of the ejecta are all listed in the online catalogue.
Meanwhile, the detail magnetic field and plasma parameters and figures of these ICMEs could also be found at the online catalogue. 
Based on this catalogue, the annual numbers of the ICMEs, the annual numbers and the ratios of MCs and shocks driven by the ICMEs, 
the magnetic field and plasma properties of the ejecta and the sheath regions are also discussed. 
The main results we obtained are:

\begin{enumerate}
\item The annual ICME numbers are well correlated with the sunspot numbers. In addition, the shock driven the ICMEs and the percentages of the ICMEs with shocks are all well correlated with the sunspot number.  

\item The numbers of the MCs did not show any correlation of the sunspot numbers.  
But, the MC percentages of the ICMEs show obvious anti-correlation with sunspot numbers. 
This confirm the previous result that we can observe the MC with higher possibilities in solar minimum\citep[e.g.][and reference therein]{2006SoPh..239..449W,Wu2010}.

\item The distribution of the magnetic field and plasma parameters ($B$, $B_s$, $v$, $v_xB_s$, $T_p$ and $N_p$) of ICME are discussed. 
In addition, by compare the parameters of the MC and the Non-MC ICMEs, we confirm the result that the MCs are stronger than the Non-MC ICMEs. 

\item The yearly variation of these parameters are discussed. It is found that almost all these parameters, expected the proton number density, are varied with the solar cycle. Meanwhile, the mean and maximum values of all the parameters after 2008 are smaller than before.

\item By compared the parameters of ICMEs with and without shocks, we found that the ICMEs with shocks are much stronger than the ICMEs without shocks. 
Thus, a strong ICME could be expected if we observed a shock ahead of it. 
In addition, the distributions of the parameters in the sheath regions have also been discussed. 
We found that the magnetic field and the velocity are higher in the sheath regions than them in the ejecta of ICMEs.
\end{enumerate}

\acknowledgments{ We acknowledge the use of the data from WIND and ACE spacecraft.
This work is supported by grants from MOST 973 key
project (2011CB811403), CAS (Key Research Program KZZD-EW-01 and 100-Talent Program),
NSFC (41131065, 41121003, 41274173, 41222031 and 41404134), the fundamental
research funds for the central universities and the Specialized Research Fund for State Key Laboratories.}


\begin{thebibliography}{33}
\providecommand{\natexlab}[1]{#1}
\expandafter\ifx\csname urlstyle\endcsname\relax
  \providecommand{\doi}[1]{doi:\discretionary{}{}{}#1}\else
  \providecommand{\doi}{doi:\discretionary{}{}{}\begingroup
  \urlstyle{rm}\Url}\fi

\bibitem[{\textit{Burlaga et~al.}(1981)\textit{Burlaga, Sittler, Mariani, and
  Schwenn}}]{Burlaga1981}
Burlaga, L., E.~Sittler, F.~Mariani, and R.~Schwenn, {Magnetic loop behind an
  interplanetary shock - Voyager, Helios, and IMP 8 observations},
  \textit{Journal of Geophysical Research: Space Physics}, \textit{86},
  6673--6684, \doi{10.1029/JA086iA08p06673}, 1981.

\bibitem[{\textit{Burlaga et~al.}(2001)\textit{Burlaga, Skoug, Smith, Webb,
  Zurbuchen, and Reinard}}]{2001JGR...10620957B}
Burlaga, L.~F., R.~M. Skoug, C.~W. Smith, D.~F. Webb, T.~H. Zurbuchen, and
  A.~Reinard, {Fast ejecta during the ascending phase of solar cycle 23: ACE
  observations, 1998-1999}, \textit{Journal of Geophysical Research},
  \textit{106}(A), 20,957--20,978, 2001.

\bibitem[{\textit{Cane and Lario}(2006)}]{Cane:2006kg}
Cane, H.~V., and D.~Lario, {An Introduction to CMEs and Energetic Particles},
  \textit{Space Science Reviews}, \textit{123}(1-3), 45--56, 2006.

\bibitem[{\textit{Cane and Richardson}(2003)}]{Cane2003d}
Cane, H.~V., and I.~G. Richardson, {Interplanetary coronal mass ejections in
  the near-Earth solar wind during 1996-2002}, \textit{Journal of Geophysical
  Research: Space Physics}, \textit{108}, \doi{10.1029/2002JA009817}, 2003.

\bibitem[{\textit{Cremades and Bothmer}(2004)}]{Cremades:2004cz}
Cremades, H., and V.~Bothmer, {On the three-dimensional configuration of
  coronal mass ejections}, \textit{Astronomy and Astrophysics},
  \textit{422}(1), 307--322, 2004.

\bibitem[{\textit{Gonzalez et~al.}(1994)\textit{Gonzalez, Joselyn, Kamide,
  Kroehl, Rostoker, Tsurutani, and Vasyliunas}}]{Gonzalez1994}
Gonzalez, W.~D., J.~A. Joselyn, Y.~Kamide, H.~W. Kroehl, G.~Rostoker, B.~T.
  Tsurutani, and V.~M. Vasyliunas, {What is a geomagnetic storm?},
  \textit{Journal of Geophysical Research}, \textit{99}(A4), 5771--5792, 1994.

\bibitem[{\textit{Gopalswamy}(2006)}]{2006SSRv..124..145G}
Gopalswamy, N., {Properties of Interplanetary Coronal Mass Ejections},
  \textit{Space Science Reviews}, \textit{124}(1), 145--168, 2006.

\bibitem[{\textit{Gopalswamy et~al.}(2000)\textit{Gopalswamy, Lara, Lepping,
  Kaiser, Berdichevsky, and {St Cyr}}}]{Gopalswamy2000}
Gopalswamy, N., A.~Lara, R.~P. Lepping, M.~L. Kaiser, D.~Berdichevsky, and
  O.~C. {St Cyr}, {Interplanetary acceleration of coronal mass ejections},
  \textit{Geophysical Research Letters}, \textit{27}, 145, 2000.

\bibitem[{\textit{Gopalswamy et~al.}(2003)\textit{Gopalswamy, Lara, Yashiro,
  Nunes, and Howard}}]{2003ESASP.535..403G}
Gopalswamy, N., A.~Lara, S.~Yashiro, S.~Nunes, and R.~a. Howard, {Coronal mass
  ejection activity during solar cycle 23}, \textit{In: Solar variability as an
  input to the Earth's environment. International Solar Cycle Studies (ISCS)
  Symposium}, \textit{535}, 403--414, 2003.

\bibitem[{\textit{Gosling et~al.}(1973)\textit{Gosling, Pizzo, and
  Bame}}]{Gosling1973}
Gosling, J.~T., V.~Pizzo, and S.~J. Bame, {Anomalously low proton temperatures
  in the solar wind following interplanetary shock waves—evidence for
  magnetic bottles?}, \textit{Journal of Geophysical Research},
  \textit{78}(13), 2001, \doi{10.1029/JA078i013p02001}, 1973.

\bibitem[{\textit{Gui et~al.}(2011)\textit{Gui, Shen, Wang, Ye, and
  Wang}}]{Gui:2011dw}
Gui, B., C.~Shen, Y.~Wang, P.~Ye, and S.~Wang, {Quantitative Analysis of CME
  Deflections in the Corona}, \textit{Solar Physics}, \textit{271}, 111--139,
  2011.

\bibitem[{\textit{Jian et~al.}(2006)\textit{Jian, Russell, Luhmann, and
  Skoug}}]{Jian2006c}
Jian, L., C.~T. Russell, J.~G. Luhmann, and R.~M. Skoug, {Properties of
  Interplanetary Coronal Mass Ejections at One AU During 1995 – 2004},
  \textit{Solar Physics}, \textit{239}(1-2), 393--436,
  \doi{10.1007/s11207-006-0133-2}, 2006.

\bibitem[{\textit{Jian et~al.}(2011)\textit{Jian, Russell, and
  Luhmann}}]{Jian2011}
Jian, L.~K., C.~T. Russell, and J.~G. Luhmann, {Comparing Solar Minimum 23/24
  with Historical Solar Wind Records at 1 AU}, \textit{Solar Physics},
  \textit{274}(1-2), 321--344, 2011.

\bibitem[{\textit{Kilpua et~al.}(2012)\textit{Kilpua, Jian, Li, Luhmann, and
  Russell}}]{Kilpua2012}
Kilpua, E. K.~J., L.~K. Jian, Y.~Li, J.~G. Luhmann, and C.~T. Russell,
  {Observations of ICMEs and ICME-like Solar Wind Structures from 2007 – 2010
  Using Near-Earth and STEREO Observations}, \textit{Solar Physics}, 2012.

\bibitem[{\textit{Kilpua et~al.}(2014)\textit{Kilpua, Mierla, Zhukov,
  Rodriguez, Vourlidas, and Wood}}]{Kilpua2014a}
Kilpua, E. K.~J., M.~Mierla, a.~N. Zhukov, L.~Rodriguez, a.~Vourlidas, and
  B.~Wood, {Solar Sources of Interplanetary Coronal Mass Ejections During the
  Solar Cycle 23/24 Minimum}, \textit{Solar Physics}, \textit{289}(10),
  3773--3797, \doi{10.1007/s11207-014-0552-4}, 2014.

\bibitem[{\textit{Kilpua et~al.}(2009)}]{Kilpua2009}
Kilpua, E. K.~J., et~al., {Multispacecraft Observations of Magnetic Clouds and
  Their Solar Origins between 19 and 23 May 2007}, \textit{Solar Physics},
  \textit{254}(2), 325--344, 2009.

\bibitem[{\textit{Lepping et~al.}(2006)\textit{Lepping, Berdichevsky, Wu,
  Szabo, Narock, Mariani, Lazarus, and Quivers}}]{Lepping2006}
Lepping, R.~P., D.~B. Berdichevsky, C.-C. Wu, a.~Szabo, T.~Narock, F.~Mariani,
  a.~J. Lazarus, and a.~J. Quivers, {A summary of WIND magnetic clouds for
  years 1995-2003: model-fitted parameters, associated errors and
  classifications}, \textit{Annales Geophysicae}, \textit{24}(1), 215--245,
  \doi{10.5194/angeo-24-215-2006}, 2006.

\bibitem[{\textit{Lepping et~al.}(2014)\textit{Lepping, Wu, and
  Berdichevsky}}]{Lepping2014}
Lepping, R.~P., C.-C. Wu, and D.~B. Berdichevsky, {Yearly Comparison of
  Magnetic Cloud Parameters, Sunspot Number, and Interplanetary Quantities for
  the First 18 Years of the Wind Mission}, \textit{Solar Physics},
  \textit{290}(2), 553--578, \doi{10.1007/s11207-014-0622-7}, 2014.

\bibitem[{\textit{Lugaz and Kintner}(2012)}]{Lugaz:2012es}
Lugaz, N., and P.~Kintner, {Effect of Solar Wind Drag on the Determination of
  the Properties of Coronal Mass Ejections from Heliospheric Images},
  \textit{Solar Physics}, 2012.

\bibitem[{\textit{Richardson and Cane}(2004)}]{Richardson2004a}
Richardson, I.~G., and H.~V. Cane, {The fraction of interplanetary coronal mass
  ejections that are magnetic clouds: Evidence for a solar cycle variation},
  \textit{Geophysical Research Letters}, \textit{31}(18), 8--11,
  \doi{10.1029/2004GL020958}, 2004.

\bibitem[{\textit{Richardson and Cane}(2010)}]{Richardson:2010jq}
Richardson, I.~G., and H.~V. Cane, {Near-Earth Interplanetary Coronal Mass
  Ejections During Solar Cycle 23 (1996 - 2009): Catalog and Summary of
  Properties}, \textit{Solar Physics}, \textit{264}, 189, 2010.

\bibitem[{\textit{Shen et~al.}(2007)\textit{Shen, Wang, Ye, Zhao, Gui, and
  Wang}}]{Shen:2007ww}
Shen, C., Y.~Wang, P.~Ye, X.~P. Zhao, B.~Gui, and S.~Wang, {Strength of coronal
  mass ejection-driven shocks near the sun and their importance in predicting
  solar energetic particle events}, \textit{Astrophysical Journal},
  \textit{670}(1), 849--856, 2007.

\bibitem[{\textit{Shen et~al.}(2011)\textit{Shen, Wang, Gui, Ye, and
  Wang}}]{Shen2011}
Shen, C., Y.~Wang, B.~Gui, P.~Ye, and S.~Wang, {Kinematic Evolution of a Slow
  CME in Corona Viewed by STEREO-B on 8 October 2007}, \textit{Solar Physics},
  \textit{269}(2), 389--400, 2011.

\bibitem[{\textit{Shen et~al.}(2014)\textit{Shen, Wang, Pan, Miao, Ye, and
  Wang}}]{Shen2014a}
Shen, C., Y.~Wang, Z.~Pan, B.~Miao, P.~Ye, and S.~Wang, {Full-halo coronal mass
  ejections : Arrival at the Earth}, \textit{Journal of Geophysical Research :
  Space Physics}, p. DOI:10.1002/2014JA020001,
  \doi{10.1002/2014JA020001.Received}, 2014.

\bibitem[{\textit{Vr\v{s}nak et~al.}(2013)}]{Vrsnak2012}
Vr\v{s}nak, B., et~al., {Propagation of Interplanetary Coronal Mass Ejections:
  The Drag-Based Model}, \textit{Solar Physics}, \textit{285}(1-2), 295--315,
  2013.

\bibitem[{\textit{Wang et~al.}(2014)\textit{Wang, Wang, Shen, Shen, and
  Lugaz}}]{Wang2014b}
Wang, Y., B.~Wang, C.~Shen, F.~Shen, and N.~Lugaz, {Deflected propagation of a
  coronal mass ejection from the corona to interplanetary space},
  \textit{Journal of Geophysical Research: Space Physics}, \textit{119}, 1--16,
  \doi{10.1002/2013JA019537}, 2014.

\bibitem[{\textit{Wang et~al.}(2015)\textit{Wang, Zhou, Shen, Liu, and
  Wang}}]{Wang2015}
Wang, Y., Z.~Zhou, C.~Shen, R.~Liu, and S.~Wang, {Investigating plasma motion
  of magnetic clouds at 1 AU through a velocity-modified cylindrical force-free
  flux rope model}, \textit{Journal of Geophysical Research : Space Physics},
  \doi{10.1002/2014JA020494.Received}, 2015.

\bibitem[{\textit{Wang and Colaninno}(2014)}]{Wang2014}
Wang, Y.-M., and R.~Colaninno, {Is Solar Cycle 24 Producing More Coronal Mass
  Ejections Than Cycle 23?}, \textit{The Astrophysical Journal},
  \textit{784}(2), L27, \doi{10.1088/2041-8205/784/2/L27}, 2014.

\bibitem[{\textit{Wang et~al.}(2003)\textit{Wang, Ye, and Wang}}]{Wang2003}
Wang, Y.~M., P.~Z. Ye, and S.~Wang, {Multiple magnetic clouds: Several examples
  during March-April 2001}, \textit{Journal of Geophysical Research: Space
  Physics}, \textit{108}(A10), 1370, \doi{10.1029/2003JA009850}, 2003.

\bibitem[{\textit{Wimmer-Schweingruber et~al.}(2006)}]{2006SSRv..123..177W}
Wimmer-Schweingruber, R.~F., et~al., {Understanding Interplanetary Coronal Mass
  Ejection Signatures. Report of Working Group B}, \textit{Space Science
  Reviews}, \textit{123}(1), 177--216, 2006.

\bibitem[{\textit{Wu and Lepping}(2010)}]{Wu2010}
Wu, C.-C., and R.~P. Lepping, {Statistical Comparison of Magnetic Clouds with
  Interplanetary Coronal Mass Ejections for Solar Cycle 23}, \textit{Solar
  Physics}, p. 242, 2010.

\bibitem[{\textit{Wu and Lepping}(2015)}]{Wu2015}
Wu, C.-C., and R.~P. Lepping, {Comparisons of Characteristics of Magnetic
  Clouds and Cloud-Like Structures During 1995 – 2012}, \textit{Solar
  Physics}, \doi{10.1007/s11207-015-0656-5}, 2015.

\bibitem[{\textit{Wu et~al.}(2006)\textit{Wu, Lepping, and
  Gopalswamy}}]{2006SoPh..239..449W}
Wu, C.-C., R.~P. Lepping, and N.~Gopalswamy, {Relationships Among Magnetic
  Clouds, CMES, and Geomagnetic Storms}, \textit{Solar Physics},
  \textit{239}(1), 449--460, 2006.

\end{thebibliography}
\end{document}